\documentclass{amsart}
\usepackage[dvips]{epsfig}
\usepackage{graphics}
\usepackage{latexsym}
\usepackage{amsmath}
\usepackage{amsthm}
\usepackage{amssymb}
\usepackage{hyperref}

\newtheorem{thm}{Theorem}%[section]
\newtheorem{lem}[thm]{Lemma}

\newtheorem{cor}[thm]{Corollary}
\newtheorem{defi}[thm]{Definition}

\newtheorem{rk}[thm]{Remark}

\newenvironment{preuve}{\noindent {\it Proof}}{\hfill$\square$}

\newcommand{\rr}{{\mathbb{R}}}

\newcommand{\rd}{{\rr^d}}

\newcommand{\e}{{\varepsilon}}
\newcommand{\vip}{\vskip.2cm}
\newcommand{\E}{\mathbb{E}}
\newcommand{\cP}{{\mathcal P}}
\newcommand{\loi}{\mathcal{ L}}
\newcommand{\ds}{\displaystyle}
\newcommand{\intot}{\ds \int _0^t }
\newcommand{\intrd}{\ds \int_\rd}
\newcommand{\ah}{a^\frac{1}{2}}
\newcommand{\sq}{{^\frac{1}{2}}}
\newcommand{\cW}{{{\mathcal W}}}
\newcommand{\Wd}{{{\mathbf W}_d}}
\newcommand{\rs}{{\rho_N(s)}}
\newcommand{\ala}{\nonumber \\}
\newcommand{\indiq}{{\bf 1}}
\begin{document}

\title
{Particle approximation of some Landau equations}

\author[Nicolas Fournier]{Nicolas Fournier$^1$}

\footnotetext[1]{LAMA UMR 8050,
Facult\'e de Sciences et Technologies,
Universit\'e Paris Est, 61, avenue du G\'en\'eral de Gaulle, 94010 Cr\'eteil 
Cedex, France, {\tt nicolas.fournier@univ-paris12.fr}}

\def\shortauthorname{Nicolas Fournier}

\def\abstractname{Abstract}

\begin{abstract}
We consider a class of nonlinear partial-differential equations,
including the spatially homogeneous 
Fokker-Planck-Landau equation for Maxwell (or pseudo-Maxwell) molecules.
%A probabilistic interpretation of such a P.D.E. has been proposed in
%\cite{f,g,fgm}, in terms of some white noise-driven nonlinear stochastic 
%differential equation, and a numerical scheme based on a system of
%$n$ particles driven by $n^2$ Brownian motions has been studied.
%
%\noindent 
Continuing the work of \cite{f,g,fgm}, 
we propose a probabilistic interpretation of such a
P.D.E. in terms of a
nonlinear stochastic differential equation  driven by a standard 
Brownian motion. We derive a numerical scheme, based on a system of
$n$ particles driven by $n$ Brownian motions,
and study its rate of convergence.
%
%\noindent 
We finally deal with the possible extension of our numerical
scheme to the case of the Landau equation for soft potentials, and give
some numerical results.
\end{abstract}

\maketitle

\smallskip

\smallskip

\section{Introduction and main results}

\subsection{The equation}

Let $S_d$ be the set of symmetric $d\times d$ matrices with
real entries, and $S_d^+$ its subset of nonnegative matrices.
For $a:\rd\mapsto S_d^+$, we consider the 
partial differential equation
\begin{equation}\label{eqa}
\partial_t f_t(x) =\frac{1}{2}\sum_{i,j=1}^d
\partial_i \left\{
\int_{\rd}a_{ij}( x-y) \Big[ f_t( y) \partial_j f_t( x) -f_t( x) 
\partial_j f_t( y) \Big]dy \right\},
\end{equation}
where $\partial_t =\frac{\partial }{\partial t}$, $\partial_i =
\frac{\partial }{\partial x_i}$ and where the unknown 
$(f_t)_{t\geq 0}$ is a family of probability
density functions $(f_t)_{t\geq 0}$ on $\rd$. The spatially
homogeneous Landau
(or Fokker-Planck-Landau) equation
corresponds, in dimension $d\geq 2$,  to the case where for some
$\kappa:\rr_+\mapsto \rr_+$,
\begin{equation} \label{aLandau}
a_{ij}(z)=\kappa(|z|^2) (|z|^{2}\delta_{ij}-z_iz_j).
\end{equation}
Physically, one assumes that $\kappa(r)=r^{\gamma/2}$, for some 
$\gamma \in [-3,1]$.
One talks of soft potentials when $\gamma<0$, Maxwell molecules when
$\gamma=0$, and hard potentials when $\gamma>0$. We consider in this
paper the case of Maxwell molecules, or of pseudo-Maxwell molecules,
where $\kappa$ is supposed to be smooth and bounded.

\vip

This equation arises as a limit of the Boltzmann equation when
all the collisions become grazing.
We refer to Villani \cite{v,v2,v3} and the many references therein
for physical and mathematical details on this topic.
See Cordier-Mancini \cite{cm} and Buet-Cordier-Filbet \cite{bcf} 
for a review on deterministic numerical
methods to solve (\ref{eqa}).

\subsection{Notation} 
Let $\cP=\cP(\rd)$ be the set of probability measures on $\rd$,
and $\cP_k=\{\mu\in\cP, m_k(\mu)<\infty\}$, where $m_k(\mu)=\int |x|^k\mu(dx)$.

For $x,y\in \rd$, we set $|x|=(\sum_1^d x_i^2)\sq$, and $(x,y)=x^*y
=\sum_1^d x_iy_i$.
We consider the norm  $|M|=\sup \{|(Mx,x)|, |x|=1\}
=\max\{|\lambda|, \lambda$ eighenvalue of $M\}$ on $S_d$.
Recall that for $A \in S_d^+$, $\inf \{(Ax,x), |x|=1 \}=1/|A^{-1}|$. All
$A\in S_d^+$ admits a unique square root $A\sq \in S_d^+$, and 
we have $|A\sq|=|A|\sq$.

\begin{defi}
Consider $a:\rd \mapsto S_d^+$. Let $b:\rd\mapsto \rd$ be defined  by
$b_i(x)=\sum_{j=1}^d \partial_j a_{ij}(x)$.
Assume that $|a(x)|+|b(x)|\leq C(1+|x|^2)$ (which is the case
when $a$ is defined by (\ref{aLandau}) with $\kappa\in C^1_b$).
A measurable family $(P_t)_{t \geq 0}\subset \cP_2$ is said to be a
weak solution to (\ref{eqa}) if for all $t\geq 0$, 
$\sup_{[0,t]} m_2(P_s) <\infty$ and for all $\varphi \in C^2_b(\rd)$,
\begin{equation}\label{LW}
\intrd \varphi(x)P_t(dx) = \intrd \varphi(x)P_0(dx)  +
\intot ds \intrd \intrd P_s(dx)P_s(dy) L\varphi(x,y),
\end{equation}
where $L\varphi(x,y)= \frac{1}{2}\sum_{i,j=1}^d a_{ij}(x-y)\partial^2_{ij}
\varphi(x) + \sum_{i=1}^d b_i(x-y)\partial_i\varphi(x)$.
\end{defi}

All the terms make sense due to our conditions on $a$, $b$, $P_t$.
See Villani \cite{v2} for a similar formulation.

\subsection{Known results}

To our knowledge, the first (and only) paper proving a rate of convergence
for a numerical
scheme to solve (\ref{eqa}) 
is that of Fontbona-Gu\'erin-M\'el\'eard \cite{fgm}. 
Their method relies on a stochastic particle system.
The aim of this paper is to go further in this direction.

Let us thus recall briefly the method
of \cite{fgm}, relying on the probabilistic interpretation of (\ref{eqa}) 
developped by Funaki \cite{f}, Gu\'erin \cite{g}.

Let $\sigma:\rd \mapsto S_d^+$ and $b:\rd\mapsto\rd$ be Lipschitz continuous
functions, and let $P_0\in\cP_2$. A $\rd$-valued process $(X_t)_{t\geq 0}$
is said to
solve $E_0(P_0,\sigma,b)$ if $\loi(X_0)=P_0$, and if for all $t\geq 0$,
setting $P_t=\loi(X_t)$,
\begin{equation}\label{sdew}
X_t=X_0+\intot\intrd \sigma(X_s-x)W_P(dx,ds) + \intot \intrd b(X_s-x)P_s(dx)ds.
\end{equation}
Here $W_P(dx,dt)$ is a $\rd$-valued white noise on $[0,\infty)\times \rd$,
independent of $X_0$, 
with independent coordinates, each of which having covariance measure
$P_t(dx)dt$ (see Walsh \cite{w}). 

Existence and uniqueness in law for $E_0(P_0,\sigma,b)$ have been proved in
Gu\'erin \cite{g}. If furthermore $\sigma(x)\sigma^*(x)=a(x)$
and $b_i(x)=\sum_{j=1}^d \partial_ja_{ij}(x)$, then $(P_t)_{t\geq 0}$
is a weak solution to (\ref{eqa}). The condition that 
$\sigma$ and $b$ are Lipschitz continuous is satisfied in the case
of the Landau equation for Maxwell or pseudo-Maxwell molecules.

\vip

In \cite{fgm}, one considers an exchangeable 
stochastic particle system 
$(X^{i,n}_t)_{t\geq 0,i=1,\dots,n}$, satisfying a S.D.E. driven by $n^2$ Brownian
motions. It is then shown that
one may find a coupling between a solution $(X_t^1)_{t\geq 0}$ 
to $E_0(P_0,\sigma,b)$ and such a
particle system in such a way that
\begin{equation*}%\label{fgmr1}
\E\left[ \sup_{[0,T]} |X^{1,n}_t - X_t^1|^2\right] 
\leq C_{T} n^{-2/(d+4)},
\end{equation*}
under the condition
that $P_0$ has a finite moment of order $d+5$.
The proof relies on a clever coupling 
between the the white noise and $n$ Brownian motions.
In particular, one has to assume that $P_t$ has a density for all $t>0$,
in order to guarantee the uniqueness of some optimal couplings.

\subsection{Another approach}
For $a:\rd\mapsto S_d^+$, $b:\rd\mapsto\rd$ 
satisfying $|a(x)|+|b(x)|\leq C (1+|x|^2)$
and $\mu \in \cP_2(\rd)$, we introduce
\begin{equation*}
a(x,\mu)= \int_\rd a(x-y)\mu(dy), \quad b(x,\mu)= \int_\rd b(x-y)\mu(dy).
\end{equation*}
For each $x \in \rd$, $\mu\in \cP_2$, 
$a(x,\mu)$ is a nonnegative symmetric matrix
and thus admits an unique symmetric nonnegative square root $\ah(x,\mu)
:= [a(x,\mu)]^{\frac{1}{2}}$. 

\vip

Denote by $\Wd$ the law of the $d$-dimensional Brownian motion, consider
$P_0\in\cP_2$, and let $(X_0,B)\sim P_0 \otimes \Wd$.
We say that a $\rd$-valued process $(X_t)_{t\geq 0}$
solves $E_1(P_0,a,b)$ (or $E_1(P_0,a,b,X_0,B)$ when needed) 
if $\E[\sup_{[0,T]} |X_t|^2]<\infty$ for all $T$ and
if for all $t\geq 0$, setting $P_t=\loi(X_t)$,
\begin{equation}\label{sdeb}
X_t=X_0+\intot \ah(X_s,P_s)dB_s + \intot b(X_s,P_s)ds.
\end{equation}

This equation is nonlinear in the sense that its coefficients
involve the law of the solution.
Compared to (\ref{sdew}), equation (\ref{sdeb}) 
is simpler, since it is driven by
a finite-dimensional Brownian motion, and since the nonlinearity does
not involve the driving process.
However, one may check that at least formally, solutions
to (\ref{sdew}) and (\ref{sdeb}) have the same law.
The link with (\ref{eqa}) relies on a simple application of the It\^o formula.

\begin{rk}\label{lien}
Let $(X_t)_{t\geq 0}$ solve $E_1(P_0,a,b)$. Assume that 
$b_i=\sum_{j=1}^d \partial_j a_{ij}$, and that
$|a(x)|+|b(x)|\leq C(1+|x|^2)$. Then $(P_t)_{t\geq 0}:=(\loi(X_t))_{t\geq 0}$ 
is a weak solution to (\ref{eqa}).
\end{rk}

\vip

The natural linearization of (\ref{sdeb})
consists of considering $n$ particles $(X^{i,n}_t)_{t\geq 0, i=1,\dots,n}$
solving
\begin{equation}\label{sden}
X_t^{i,n}=X_0^i+  \intot \ah\left(X_s^{i,n},
\frac{1}{n}\sum_1^n\delta_{X^{k,n}_s}\right) 
dB^{i}_s + \intot b\left(X_s^{i,n},\frac{1}{n}\sum_1^n\delta_{X^{k,n}_s} \right)ds.
\end{equation}
Here $(X_0^i,B^i)_{i=1,\dots,n}$ are i.i.d. with law $P_0\otimes \Wd$.
We thus use $n$ Brownian motions. When linearizing (\ref{sdew}), 
one needs to use $n^2$ Brownian motions, since the white noise
is infinite dimensional.

However, one may check that the solution to (\ref{sden}) and the particle
system built in \cite{fgm} have the same distribution (provided
$\sigma\sigma^*=a$ in \cite[Equation (4)]{fgm}).

\subsection{Main results} 

The main result of this paper is the following.

\begin{thm}\label{main}
Assume that $b$ is Lipschitz continuous, that $a$ is of class $C^2$, 
with all its
derivatives of order $2$ bounded, and that $P_0\in\cP_2$.

(i) There is strong existence and uniqueness for $E_1(P_0,a,b)$:
for any $(X_0,B)\sim P_0\otimes\Wd$, 
there is an unique solution $(X_t)_{t\geq 0}$ to 
$E_1(P_0,a,b,X_0,B)$.

(ii) Let $(X_0^i,B^i)_{i=1,\dots,n}$ be i.i.d. with law $P_0\otimes\Wd$.
There is an unique solution $(X^{i,n}_t)_{t\geq 0,i=1,\dots,n}$ to (\ref{sden}). 
Assume that $P_0\in\cP_4$, and consider the unique solution 
$(X^1_t)_{t\geq 0}$ to $E_1(P_0,a,b,X_0^1,B^1)$. 
There is a constant $C_T$ depending only on $d,P_0,a,b,T$ such that
\begin{equation}\label{obj1}
\E\left[ \sup_{[0,T]} |X^{1,n}_t - X^1_t|^2\right] 
\leq C_{T} \int_0^T  
\min\left( n^{-1/2} , n^{-1} \sup_{x\in\rd} (1+|a(x,P_t)^{-1}|) \right) dt
\leq C_T n^{-1/2}.
\end{equation}
\end{thm}

In the general case, we thus prove a rate of convergence in $n^{-1/2}$, 
which is faster than $n^{-2/(d+4)}$. If we have
some information on the nondegeneracy of $a(x,P_t)$, then $\ah(x,\mu)$ is
smooth around $\mu \simeq P_s$, and we can get a better rate
of convergence.

Assume for example that $a$ is uniformly elliptic (which is unfortunately
not the case of (\ref{aLandau}), since $a(x)x=0$ for all $x\in \rd$). 
Then
$\sup_x |a(x,P_t)^{-1}|\leq \sup_y |a(y)^{-1}| <\infty$, and we get
a convergence rate in $n^{-1}$.

In the case of the Landau equation for true
Maxwell molecules, we obtain the following result.

\begin{cor}\label{cormax}
Consider the Landau equation for Maxwell molecules, where $a$
is given by (\ref{aLandau}) with $\kappa\equiv 1$ and 
$b_i(x)=\sum_{j=1}^d \partial_ja_{ij}(x)=-(d-1)x_i$.
Then $a,b$ satisfy the assumptions of Theorem \ref{main}.
Let $P_0 \in \cP_4$, and adopt the notation of Theorem \ref{main}-(ii).

(i) We have $\E[ \sup_{[0,T]} |X^{1,n}_t - X^1_t|^2] 
\leq C_{T} n^{-1} (1+\log n)$.

(ii) Set $x_0=\int x P_0(dx)$. If $a(x_0,P_0)$ is invertible, then  
$\E[ \sup_{[0,T]} |X^{1,n}_t - X^1_t|^2] \leq C_{T} n^{-1}$.
\end{cor}

We finally consider the case of pseudo-Maxwell molecules. 

\begin{cor}\label{corpm}
Consider the Landau equation for pseudo-Maxwell molecules, where $a$
is given by (\ref{aLandau}) with $\kappa \in C^2(\rr_+)$, and
$b_i(x)=\sum_{j=1}^d \partial_ja_{ij}(x)=-(d-1)\kappa(|x|^2)x_i$. Assume that
$\kappa'$ has a bounded support.
Then $a,b$ satisfy the assumptions of Theorem \ref{main}-(ii).

Assume furthermore that $P_0 \in \cP_4$ has a density with
a finite entropy 
$\int P_0(x)\log P_0(x) dx <\infty$, and that $\kappa$ is bounded below
by a positive constant.
With the notation of Theorem \ref{main}, we have
$\E[ \sup_{[0,T]} |X^{1,n}_t - X^1_t|^2] 
\leq C_{T} n^{-1}$.
\end{cor}

\subsection{Time discretization}
To get a simulable particle system, it remains to discretize time
in (\ref{sden}). Let $N \geq 1$, and consider  
$\rho_N(s)= \sum_{k\geq 0} \frac{k}{N}\indiq_{s\in[k/N,(k+1)/N)}$.
Consider the simulable particle system 
$(X_t^{i,n,N})_{t\geq 0, i=1,\dots,n}$ defined by
\begin{equation}\label{sdenN}
X_t^{i,n,N}=X_0^i+  
\intot \ah\left(X_\rs^{i,n,N},
\frac{1}{n}\sum_1^n\delta_{X^{k,n,N}_\rs}\right) dB^{i}_s 
+ \intot b\left(X_\rs^{i,n,N},\frac{1}{n}\sum_1^n\delta_{X^{k,n,N}_\rs} \right)ds.
\end{equation}

\begin{thm}\label{totaldisc}
Assume that $b$ is Lipschitz continuous, that $a$ is of class $C^2$, 
with all its derivatives of order $2$ bounded, and that $P_0\in\cP_2$.
Let $(X_0^i,B^i)_{i=1,\dots,n}$ be i.i.d. with law $P_0\otimes \Wd$.
Consider the unique solutions $(X^{i,n}_t)_{t\geq 0,i=1,\dots,n}$ to (\ref{sden})
and  $(X^{i,n,N}_t)_{t\geq 0,i=1,\dots,n}$ to (\ref{sdenN}). Then
there is a constant $C_T$ depending only on $d,P_0,a,b,T$ such that
\begin{equation}\label{obj4}
\E\left[ \sup_{[0,T]}|X^{1,n}_t- X^{1,n,N}_t |^2 \right] \leq C_T N^{-1}.
\end{equation}
\end{thm}

\subsection{Conclusion}

Choosing for example $a,b$, and $P_0$ as in Corollary \ref{cormax}-(ii)
or as in Corollary \ref{corpm},
denoting by $(P_t)_{t\geq 0}=(\loi(X^1_t))_{t\geq 0}$ the weak solution to the
corresponding Landau equation, we obtain
for any $\varphi \in C^1_b$, by exchangeability,
\begin{align*}
\sup_{[0,T]}\E\left[\left|\frac{1}{n}\sum_1^n \varphi(X^{i,n,N}_t) 
- \intrd \varphi(x)P_t(dx) \right| \right]\leq 
C_T ||\varphi'||_\infty \sqrt{n^{-1}+N^{-1}}.
\end{align*}
Thus if one simulates the discretized particle system (\ref{sdenN}), 
and if one computes $\frac{1}{n}\sum_1^n \varphi(X^{i,n,N}_t) $,
we get an approximation of $\int \varphi(x)P_t(dx)$, with
a reasonnable error.

\subsection{Plan of the paper}
In Section \ref{pr}, we give the proofs of Theorems \ref{main}
and \ref{totaldisc}. Section \ref{ell} is devoted to the proofs of Corollaries
\ref{cormax} and \ref{corpm}.

In Section \ref{soft}, we briefly deal with the case of soft potentials,
but our theoritical results do not extend well.
Numerical results are given in Section \ref{num}.
Finally an appendix lies at the end of the paper.

\section{General proofs}\label{pr}

In the whole section, we assume that $P_0 \in \cP_2$, that
$a:\rd\mapsto S_d^+$ is of class
$C^2$, with bounded derivatives of order two, and that $b:\rd\mapsto \rd$
is Lipschitz continuous.
We denote by $C$ (resp. $C_T$, $C_{T,p}$) a constant
which depend only on $a,b,d,P_0$ (resp. additionally  on $T$, on $T,p$)
and whose value may change from line to line.

\vip

For $\mu,\nu \in \cP_2$, we set 
$\cW^2_2(\mu,\nu) = \min \left\{ \E \left[|X-Y|^2 \right] ; 
\; \loi(X)=\mu,\loi(Y)=\nu\right\}$. See Villani \cite{v4} for many informations
on the Wasserstein distance $\cW_2$.

\subsection{Preliminaries}

Our results are mainly based on the two
following Lemmas.

\begin{lem}\label{ll}
For all $\mu,\nu \in \cP_2$, all $x,y \in \rd$, 
\begin{align*}
&|\ah(x,\mu)-\ah(y,\nu)|^2 + |b(x,\mu)-b(y,\nu)|^2 
\leq C(|x-y|^2+\cW_2^2(\mu,\nu)), \ala
&|\ah(x,\mu)|^2+ |b(x,\mu)|^2 
\leq C(1+m_2(\mu)+|x|^2).
\end{align*}
\end{lem}

\begin{proof}
{\it Step 1.} For $\mu\in \cP_2$ fixed, we consider the map $A:\rd\mapsto S_d^+$
defined by $A(x)=a(x,\mu)$. Then  
$D^2 A(x)= \int_{\rd} D^2 a(x-y)\mu(dy)$, is clearly
uniformly bounded. Lemma \ref{a1} ensures us that
$||D (A\sq)||_\infty $ is uniformly bounded, so that
$|\ah(x,\mu) - \ah(y,\mu) |=|A\sq(x)-A\sq(y)| \leq C|x-y|$.

\vip

{\it Step 2.} We now fix $x\in \rd$, and consider $\mu,\nu \in \cP_2$. 
We introduce a couple $(X,Y)$
of random variables such that $\loi(X)=\mu$, $\loi(Y)=\nu$, and 
$\cW_2^2(\mu,\nu)=
\E [|X-Y|^2]$. We define $A:\rr\mapsto S_d^+$ by
$A(t)=\E\left[a(x-[tX+(1-t)Y]) \right]$. Then
$A(0)=\E[a(x-Y)]= a(x,\nu)$ while $A(1)=\E[a(x-X)]=a(x,\mu)$. Furthermore,
$$
|D^2A(t)|= |\E[|X-Y|^2 D^2a(x-[tX+(1-t)Y]) ]| \leq ||D^2a||_\infty
\E[|X-Y|^2]= C \cW_2^2(\mu,\nu).
$$ 
Lemma \ref{a1} 
ensures us that $||(A\sq)'||_\infty \leq C \cW_2(\mu,\nu)$, so that
$|\ah(x,\mu) - \ah(x,\nu) |= |A\sq(1)-A\sq(0)|  \leq C \cW_2 (\mu,\nu)$.

\vip

{\it Step 3.} The growth estimate (for $a$) follows from the Lipschitz estimate,
since $|\ah(0,\delta_0)|^2=|\ah(0)|^2 <\infty$,
and since $\cW_2^2(\mu,\delta_0)=m_2(\mu)$.

\vip

{\it Step 4.} The case of $b$ is much simpler. For $\mu,\nu\in \cP_2$, 
we introduce $X,Y$ as in Step 2.
Then $|b(x,\mu)-b(y,\nu)|^2=|\E[b(x-X)-b(y-Y)]|^2\leq C(|x-y|^2+\E[|X-Y|]^2)
\leq  C(|x-y|^2+\cW_2^2(\mu,\nu))$.
The growth estimate follows from the Lipschitz estimate,
since $|b(0,\delta_0)|^2=|b(0)|^2 <\infty$.
\end{proof}

\begin{lem}\label{nesti}
Let $Y_i$ be i.i.d. $\rd$-valued random variables with common law $\mu\in\cP_4$.
Then
\begin{equation*}
\E\left[
\left|a(Y_1,\mu)-a\left(Y_1,\frac{1}{n}\sum_1^n\delta_{Y_i}\right) \right|^2 
+\left|b(Y_1,\mu)-b\left(Y_1,\frac{1}{n}\sum_1^n\delta_{Y_i}\right) \right|^2 
\right] \leq C \frac{1+m_4(\mu)}{n}.
\end{equation*}
\end{lem}

\begin{proof}
We denote by $\E_1$ the expectation concerning only $Y_1$, and by $\E_{2,n}$
the expectation concerning only $Y_2,\dots,Y_n$. We observe that for all
$i=2,\dots,n$, we have  $a(Y_1,\mu)=\E_{2,n}[a(Y_1-Y_i)]$, whence
$a(Y_1,\mu)=\E_{2,n}[\frac{1}{n-1}\sum_2^n a(Y_1-Y_i)]$. 
We also have $a(Y_1,\frac{1}{n}\sum_1^n\delta_{Y_i})=\frac{1}{n}
\sum_1^n a(Y_1-Y_i)$.
As a consequence,
\begin{align*}
&\E\left[\left|a\left(Y_1,\frac{1}{n}\sum_1^n\delta_{Y_i}\right)
-a(Y_1,\mu) \right|^2 \right] \leq
2\E\left[\left|\frac{1}{n}\sum_1^n a(Y_1-Y_i)- 
\frac{1}{n-1}\sum_2^{n} a(Y_1-Y_i) \right|^2 \right]\ala
&\hskip1.5cm+2\E_1\left\{ 
\E_{2,n}\left[\left|\frac{1}{n-1}\sum_2^n a(Y_1-Y_i)- 
\E_{2,n}\left[\frac{1}{n-1}\sum_2^{n} a(Y_1-Y_i)\right] \right|^2 \right]
\right\}=:2I_{n}+2J_{n}.
\end{align*}
An immediate computation, using that $|a(x)|\leq C(1+|x|^2)$, shows
that $I_n \leq C(1+m_4(\mu))/n^2$.
On the other hand, since the random 
variables $Y_1-Y_i$ are i.i.d. under $\E_{2,n}$,
\begin{align*}
J_n
\leq& \E_1 \left\{ \sum_{k,l=1}^d Var_{2,n}
\left(\frac{1}{n-1}\sum_2^n a_{kl}(Y_1-Y_i)\right)\right\}
\leq \frac{1}{n-1}  \E_1 \left\{ \sum_{k,l=1}^d Var_{2,n} a_{kl}(Y_1-Y_2)\right\}
\ala
\leq & \frac{C}{n-1} \E_1\left\{ \E_{2,n}\left[|a(Y_1-Y_2)|^2\right]\right\}
\leq \frac{C}{n} \E \left[|a(Y_1-Y_2)|^2\right] \leq  \frac{C}{n}(1+m_4(\mu)),
\end{align*}
again since $|a(x)|\leq C(1+|x|^2)$. The same computation holds
for $b$, replacing everywhere $m_4(\mu)$ by $m_2(\mu)$,
since $|b(x)|\leq C(1+|x|)$.
\end{proof}

\subsection{Convergence proofs}

We start this subsection with some
moment estimates.

\begin{lem}\label{mom}
(i) Let $(X_t)_{t\geq 0}$ solve $E_1(P_0,a,b)$. Assume that
$m_p(P_0)<\infty$
for some $p\geq 2$. Then $\E[\sup_{[0,T]}|X_t|^p]<\infty$ for all $T>0$.

(ii) Let $(X^{i,n}_t)_{t\geq 0, i=1,\dots,n}$ solve (\ref{sden}). For all
$0\leq s \leq t\leq T$, $\E[|X_t^{1,n}-X_s^{1,n}|^2] \leq C_T |t-s|$.
\end{lem}

\begin{proof}
{\it Point (i).} 
Set $P_t=\loi(X_t)$. Using the Burkholder-Davies-Gundy inequality
for the Brownian part, and the H\"older inequality for the drift part,
we obtain, for all $0\leq t \leq T$,
\begin{align*}
\E\left[\sup_{[0,t]}|X_s|^p\right]\leq C_p \E[|X_0|^p] +
C_p \intot ds \E\left[|\ah(X_s,P_s)|^p \right] + C_{p,T} \intot ds  
\E\left[|b(X_s,P_s)|^p \right].
\end{align*}
But Lemma \ref{ll} implies that $\E[|\ah(X_s,P_s)|^p+|b(X_s,P_s)|^p ]
\leq C_p \E[1+|X_s|^p+m_2(P_s)^{p/2}]$. Furthermore, since  
$P_s=\loi(X_s)$ and $p\geq 2$, we deduce that $m_2(P_s)^{p/2}\leq \E[|X_s|^p]$. 
As a conclusion, 
$\E[\sup_{[0,t]}|X_s|^p]\leq C_p \E[|X_0|^p] + C_{p,T}\int_0^t
ds \E[1+|X_s|^p]$, whence the result by the Gronwall Lemma.

\vip

{\it Point (ii).} Using the Cauchy-Scharz and Doob inequalities, 
we see that for $0\leq s \leq t \leq T$,
\begin{align}\label{cc}
E\left[ |X^{1,n}_t-X^{1,n}_s|^2\right] \leq&
C \int_s^t du \E\left[|\ah(X_u^{1,n},\frac{1}{n}\sum_{1}^n \delta_{X^{i,n}_u})|^2 
\right]
+ C_{T} \int_s^t du  
\E\left[|b(X_u^{1,n},\frac{1}{n}\sum_{1}^n \delta_{X^{i,n}_u})|^2 \right] \ala
\leq& C_T \int_s^t du \E\left[ 1+ |X^{1,n}_u|^2 
+m_2\left(\frac{1}{n}\sum_{1}^n \delta_{X^{i,n}_u} \right)\right]
\leq  C_T \int_s^t du \E\left[ 1+ |X^{1,n}_u|^2 \right].
\end{align}
We used Lemma \ref{ll} and that
$\E[m_2 (\frac{1}{n}\sum_{1}^n \delta_{X^{i,n}_u})]=\frac{1}{n}\sum_{1}^n 
\E[|X^{i,n}_u|^2  ]=\E[|X^{1,n}_u|^2  ] $ by exchangeability. 
Applying (\ref{cc}) with $s=0$, we get $E[ |X^{1,n}_t|^2] \leq  C \E[|X_0^1|^2] +
C_T\int_0^t du [1+\E[|X^{1,n}_u|^2]du$. The Gronwall Lemma allows us to conclude
that $\sup_{[0,T]}E[|X^{1,n}_t|^2] \leq C_T$.
Applying a second time (\ref{cc}), we deduce that 
$E[ |X^{1,n}_t-X^{1,n}_s|^2] \leq C_T |t-s|$.
\end{proof}

\begin{preuve} {\it of Theorem \ref{main}.} We consider $P_0\in\cP_2$ fixed.

\vip

{\it Point (i).} Let $(X_0,B)\sim P_0\otimes\Wd$.

{\it Uniqueness.} Assume that we have
two solutions $X,Y$ to $E_1(P_0,a,b,X_0,B)$, and set $P_t=\loi(X_t)$, 
$Q_t=\loi(Y_t)$. Using
the Cauchy-Schwarz and Doob inequalities, we obtain, 
for $0\leq t\leq T$,
\begin{align}\label{tech}
\E\left[\sup_{[0,t]} |X_s-Y_s|^2\right]\leq& C_T 
\intot \E[|\ah(X_s,P_s)
- \ah(Y_s,Q_s) |^2
+ |b(X_s,P_s) - b(Y_s,Q_s) |^2]ds \ala
\leq& C_T \intot \E\left[ |X_s-Y_s|^2 + \cW_2^2(P_s,Q_s)
\right]ds 
\leq  C_T \intot \E\left[ |X_s-Y_s|^2\right]ds.
\end{align}
We used Lemma \ref{ll} and the obvious inequality  
$\cW_2^2(P_s,Q_s) \leq \E[|X_s-Y_s|^2]$. The Gronwall Lemma
allows us to conclude that $X=Y$.

{\it Existence.} We consider the following Picard iteration: 
set $X_t^0=X_0$, and define, for $n\geq 0$,  $t\geq 0$,
\begin{equation}\label{pic}
X^{n+1}_t=X_0+\intot \ah(X^n_s,\loi(X^n_s))dB_s + 
\intot b(X_s^n, \loi(X^n_s))ds.
\end{equation}
We get as in (\ref{tech}), for $0\leq t \leq T$,
$\E[\sup_{[0,t]} |X_s^{n+1}-X_s^{n}|^2]\leq C_T 
\int_0^t \E[ |X_s^n-X_s^{n-1}|^2 ]ds$. Thus there classically 
exists $(X_t)_{t\geq 0}$
such that $\lim_n \E[\sup_{[0,T]} |X_t^{n}-X_t|^2]=0$ for all $T$,
which implies that $\lim_n\sup_{[0,T]}\cW_2^2(\loi(X^n_t),\loi(X_t)) =0$. 
Passing to the limit in (\ref{pic}), we see that
$X$ solves $E_1(P_0,a,b,X_0,B)$.

\vip

{\it Point (ii).} First of all, the strong existence and uniqueness
for (\ref{sden}) follows from standard theory (see e.g. Stroock-Varadhan
\cite{sv}), since 
for each $i$, the maps
$(x_1,\dots,x_n) \mapsto b(x^i,\frac{1}{n}\sum_1^n\delta_{x^k})$
and $(x_1,\dots,x_n) \mapsto \ah(x^i,\frac{1}{n}\sum_1^n\delta_{x^k})$ 
are Lipschitz continuous (use Lemmas \ref{ll} and \ref{a2}).

We now consider $(X_0^i,B^i)$  i.i.d. with law $P_0\otimes \Wd$,
the solution $(X^{i,n}_t)_{t\geq 0,i=1,\dots,n}$ to (\ref{sden}), and
for each $i=1,\dots,n$, the unique solution $(X^i_t)_{t\geq 0}$
to $E_1(P_0,a,b,X_0^i,B^i)$. 
For each $t\geq 0$, let $P_t=\loi(X^1_t)=\dots=\loi(X^n_t)$.
Due to the Cauchy-Schwarz and Doob inequalities, for $0\leq t \leq T$,
\begin{align*}
&\E\left[\sup_{[0,t]}|X^{1,n}_s-X^1_s|^2\right] 
\leq C_T \intot ds \E\Big[|\ah\left(X^{1,n}_s,\frac{1}{n}\sum_1^n
\delta_{X^{i,n}_s}\right) - \ah(X^{1}_s,P_s)|^2\ala
&\hskip6cm + |b\left(X^{1,n}_s,\frac{1}{n}\sum_1^n
\delta_{X^{i,n}_s}\right) - b(X^{1}_s,P_s)|^2 \Big] \ala
& \leq C_T \intot ds \Big( \E\Big[|\ah\left(X^{1,n}_s,\frac{1}{n}\sum_1^n
\delta_{X^{i,n}_s}\right) - \ah\left(X^{1}_s,\frac{1}{n}\sum_1^n
\delta_{X^{i}_s} \right)|^2 \ala
&\hskip4cm + |b\left(X^{1,n}_s,\frac{1}{n}\sum_1^n
\delta_{X^{i,n}_s}\right) - b\left(X^{1}_s,\frac{1}{n}\sum_1^n
\delta_{X^{i}_s} \right)|^2 \Big] + \Delta_n(s) \Big),
\end{align*}
where
\begin{align}\label{deltan}
\Delta_n(s):=& \E\left[\left|\ah\left(X^{1}_s,\frac{1}{n}\sum_1^n
\delta_{X^{i}_s}\right) - \ah\left(X^{1}_s,P_s\right)\right|^2
+ \left|b\left(X^{1}_s,\frac{1}{n}\sum_1^n
\delta_{X^{i}_s}\right) - b\left(X^{1}_s,P_s\right)\right|^2\right]\ala
=:& \Delta_n^1(s)+\Delta_n^2(s).
\end{align}
Using Lemmas \ref{ll} and \ref{a2}, we obtain, for $0\leq t \leq T$,
\begin{align*}
\E\left[\sup_{[0,t]}|X^{1,n}_s-X^1_s|^2\right] 
&\leq C_T \int_0^t  ds \left(
\E\left[|X^{1,n}_s-X^1_s|^2+ \cW^2_2\left(\frac{1}{n}\sum_1^n
\delta_{X^{i,n}_s},\frac{1}{n}\sum_1^n\delta_{X^{i}_s}\right)\right] 
+\Delta_n(s) \right)\ala
&\leq C_T \int_0^t ds
\E\left[|X^{1,n}_s-X^1_s|^2+ \frac{1}{n}\sum_1^n |X^{i,n}_s-X^{i}_s|^2\right]
 + C_T \int_0^t ds \Delta_n(s) \ala
& \leq C_T \intot ds \E\left[|X^{1,n}_s-X^1_s|^2\right] + C_T  \intot ds
\Delta_n(s)
\end{align*}
by exchangeability. The Gronwall Lemma ensures us that
\begin{align}\label{gron}
\E\left[\sup_{[0,T]}|X^{1,n}_s-X^1_s|^2\right] 
\leq C_T \int_0^T ds \Delta_n(s).
\end{align}
It remains to estimate $\Delta_n(s)$.  
The random variables $X^1_s,\dots,X^n_t$ are i.i.d. with
law $P_s$.
Thus Lemma \ref{nesti} shows
that $\Delta_n^2(s) \leq C (1+m_4(P_s))/n \leq C_T/n$ for $s\leq T$, due
to Lemma \ref{mom}-(i) and since $P_0\in \cP_4$ by assumption. 
Next, we use Lemma \ref{a1bis}-(i), 
the Cauchy-Schwarz inequality, and then Lemma
\ref{nesti}: for $s\leq T$,
\begin{align*}
\Delta_n^1(s) \leq& \E\left[\left|a\left(X^{1}_s,\frac{1}{n}\sum_1^n
\delta_{X^{i}_s}\right) - a\left(X^{1}_s,\loi(X^1_s)\right)\right|\right]
\leq C\left(\frac{1+m_4(P_s)}{n} \right)\sq \leq \frac{C_T}{\sqrt n}.
\end{align*}
But one may also use Lemma \ref{a1bis}-(ii) instead of Lemma
\ref{a1bis}-(i), and this gives, for $s\leq T$,
\begin{align*}
\Delta_n^1(s) \leq& \E\left[|a(X^1_s,P_s)^{-1}|
\left|a\left(X^{1}_s,\frac{1}{n}\sum_1^n
\delta_{X^{i}_s}\right) - a\left(X^{1}_s,\loi(X^1_s)\right)\right|^2\right]\ala
\leq& C \sup_x|a(x,P_s)^{-1}| \left(\frac{1+m_4(P_s)}{n} \right) 
\leq \frac{C_T}{n}\sup_x|a(x,P_s)^{-1}|.
\end{align*}
Thus $\Delta_n(s) \leq C_T n^{-1}+ C_T \min(n^{-1/2},
n^{-1}\sup_x|a(x,P_s)^{-1}|)$. Inserting this into (\ref{gron}),
we obtain (\ref{obj1}).
\end{preuve}

\vip

\begin{preuve} {\it of Theorem \ref{totaldisc}.}
Using Lemmas \ref{ll} and \ref{a2}, we get as usual (see (\ref{tech})), 
by exchangeability,
\begin{align*}
\E\left[\sup_{[0,t]}|X^{1,n}_s-X^{1,n,N}_s|^2 \right]
&\leq C_T\intot \E\left[|X^{1,n}_s-X^{1,n,N}_\rs|^2 + \cW_2^2\left( 
\frac{1}{n}\sum_1^n\delta_{X^{i,n}_s},\frac{1}{n}\sum_1^n\delta_{X^{i,n,N}_\rs} 
\right)\right] ds \ala
&\leq  C_T\intot \E\left[|X^{1,n}_s-X^{1,n,N}_\rs|^2 +
\frac{1}{n}\sum_1^n|X^{i,n}_s-X^{i,n,N}_\rs|^2 \right] ds\ala
&\leq  C_T\intot \E\left[|X^{1,n}_s-X^{1,n,N}_\rs|^2\right] ds\ala
&\leq  C_T\intot \E\left[|X^{1,n}_s-X^{1,n,N}_s|^2\right] ds
+  C_T\intot \E\left[|X^{1,n}_s-X^{1,n}_\rs|^2\right] ds.
\end{align*}
Using finally Lemma \ref{mom}-(ii), and since  
$|s-\rs|\leq 1/N$, we deduce that  
$\E [|X^{1,n}_s-X^{1,n}_\rs|^2] \leq C_T/N$. 
The Gronwall Lemma allows us to conclude.
\end{preuve}

\section{Ellipticity estimates}\label{ell}

We start with the 

\vip

\begin{preuve} {\it of Corollary \ref{cormax}.}
Recall here that $a$ is given by (\ref{aLandau}) with $\kappa\equiv 1$ and
$b(z)=-(d-1)z$. Thus $b$ is Lipschitz continuous, and the second
derivatives of $a$ are clearly bounded.
We consider a weak solution $(P_t)_{t\geq 0}$ to (\ref{eqa}).

\vip

Simple computations using (\ref{LW}) (with $\varphi(x)=x_i$,
$\varphi(x)=|x|^2$) show that
$\partial_t \int x P_t(dx)=0$ and $\partial_t m_2(P_t)=0$.
We classically may assume without loss of generality that 
$\int x P_t(dx)=\int x P_0(dx)=0$.
We also assume that $m_2(P_t)=m_2(P_0)>0$ (else 
$X_t^1=X^{1,n}_t=0$ a.s.).

\vip

We now bound from below
$(a(x,P_t)y,y)$ for $x,y \in \rd$, $t\geq 0$.

A simple computation, using that $\int x P_t(dx)=0$,
shows that $a(x,P_t)=a(x)+a(0,P_t)$. 
Thus for all $t\geq 0$, $x,y\in \rd$, 
setting $m_2^{ij}(P_t)=\int x_ix_j P_t(dx)$
$$
(a(x,P_t)y,y)\geq (a(0,P_t)y,y)=\sum_{i,j}y_iy_j 
[m_2(P_t)\delta_{ij} -m_2^{ij}(P_t)]= m_2(P_0)|y|^2
-\sum_{i,j}y_iy_jm_2^{ij}(P_t).
$$ 

Using (\ref{LW}) with $\varphi(x)=x_ix_j$, we deduce that
$$
\partial_t m_2^{ij}(P_t)= 2m_2(P_t)\delta_{ij} -2d m_2^{ij}(P_t)=
2m_2(P_0)\delta_{ij} -2d m_2^{ij}(P_t).
$$
We thus obtain
\begin{align*}
\partial_t (a(0,P_t)y,y)=& - \sum_{i,j}y_iy_j 
\partial_t m_2^{ij}(P_t) = 2 d \sum_{i,j}y_iy_j m_2^{ij}(P_t) - 2m_2(P_0)|y|^2\ala
=& 2(d-1)m_2(P_0)|y|^2 - 2d  (a(0,P_t)y,y).
\end{align*}
Set $\lambda_0=\inf\{(a(0,P_0)y,y), |y|=1\}\geq 0$
and $\lambda_1=\frac{d-1}{d}m_2(P_0)>0$. For all $t\geq 0$, all $x,y\in \rd$,
\begin{align}\label{solex} 
(a(x,P_t)y,y) \geq & (a(0,P_t)y,y)=
(a(0,P_0)y,y)e^{-2dt}+\lambda_1 |y|^2(1-e^{-2dt}) \ala
\geq& |y|^2 [\lambda_0 e^{-2dt} + \lambda_1(1-e^{-2dt})].
\end{align}

We now prove point (i). We deduce from (\ref{solex}) that 
\begin{align*}
(a(x,P_t)y,y)  \geq \lambda_1(1-e^{-2dt})|y|^2.
\end{align*}
As a consequence, $|a(x,P_t)^{-1}| \leq 1/[\lambda_1 (1-e^{-2dt})]\leq C/t+C$. 
Inserting this into (\ref{obj1}),
we get $\E[\sup_{[0,T]}|X^1_t-X^{1,n}_t|^2]\leq C_T\int_0^{T} 
\min(n^{-1/2},n^{-1}+(nt)^{-1}) dt \leq C_T n^{-1} (1+\log n)$.

\vip

To get (ii), we use (\ref{solex}) and that by assumption, 
$\lambda_0>0$. We deduce that
$$(a(x,P_t)y,y) \geq |y|^2 ( \lambda_0 e^{-2dt}+\lambda_1(1-e^{-2dt}))
\geq \min(\lambda_0,\lambda_1) |y|^2/2.$$
As a consequence, $|a(x,P_t)^{-1}|\leq 2/ \min(\lambda_0,\lambda_1)$. 
Inserting this into (\ref{obj1}),
we get $\E[\sup_{[0,T]}|X^1_t-X^{1,n}_t|^2]\leq C_T\int_0^{T} 
\min(n^{-1/2},n^{-1}) dt \leq C_T n^{-1}$.
\end{preuve}

\vip

It remains to give the

\vip

\begin{preuve} {\it of Corollary \ref{corpm}.}
Recall here that $a_{ij}(x)=\kappa(|x|^2)(|x|^2\delta_{ij}-x_ix_j)$
and that $b(x)=-(d-1)\kappa(|x|^2)x$, that $\kappa$ is $C^2$ 
and that $\kappa'$ has a bounded support,
so that $a$ has bounded derivatives of order $2$, and $b$ is Lipschitz 
continuous. 
We consider a weak solution $(P_t)_{t\geq 0}$ to (\ref{eqa}).
As previously, we classically have $m_2(P_t)=m_2(P_0)$.
Furthermore, it is again classical and widely used that the entropy
of $P_t$ is non-increasing, so that 
$\int P_t(x)\log P_t(x) dx \leq \int P_0(x)\log P_0(x) dx=C<\infty$ 
for all times, see Villani \cite{v,v2,v3}.

If we prove that there is $\lambda_0>0$ such that for all $t\geq 0$,
$x,y\in \rd$, $(a(x,P_t)y,y) \geq \lambda_0 |y|^2$, then we deduce
that $|a(x,P_t)^{-1}|$ is uniformly bounded, so that the Corollary
follows from (\ref{obj1}).

Observe that setting $\alpha_{ij}(x)=|x|^2\delta_{ij}-x_ix_j$,
we have $(a(x,P_t)y,y) \geq \lambda_1 (\alpha(x,P_t)y,y)$, where
$\lambda_1>0$ is a lowerbound of $\kappa$. But it is shown in
Desvillettes-Villani \cite[Proposition 4]{dv} that for
$E_0 \in \rr_+, H_0\in \rr_+$, there is a constant $c_{E_0,H_0}>0$ such that
for any probability density function $f$ on $\rd$ such that
$m_2(f)\leq E_0$ and $\int f(x)\log f(x) dx \leq H_0$,
$(\alpha(x,f)y,y) \geq c_{E_0,H_0} |y|^2$. Actually, they consider the case where
$\alpha_{ij}(x)=|x|^\gamma(|x|^2\delta_{ij}-x_ix_j)$ for some $\gamma>0$, but one
can check that their proof works without modification 
when $\gamma=0$.
We finally obtain $(a(x,P_t)y,y) \geq \lambda_1 c_{E_0,H_0} |y|^2$ for
all $t\geq 0$, $x,y\in\rd$, which 
concludes the proof.
\end{preuve}

\section{On soft potentials}\label{soft}

We consider in this section the spatially homogeneous Landau equation for 
soft potentials, which writes (\ref{eqa})
with
$a_{ij}(z)=|z|^\gamma(|z|^{2}\delta_{ij}-z_iz_j)$
for some $\gamma\in [-3,0)$, the Coulomb case $\gamma=-3$ being the 
most interesting from a physical point of view. 
Then we have $b_i(z)=\sum_{1}^d \partial_j a_{ij}(z)=-(d-1)|z|^\gamma z_i$.

\subsection*{Simulation with cutoff} 
We restrict our study
to the case where $\gamma \in (-2,0]$. We assume
that $P_0$ has finite moments of all orders, and has a density with
a finite entropy $\int P_0(x) \log P_0(x) dx <\infty$.

For $\e>0$ let
$\kappa_\e:\rr_+\mapsto\rr_+$ of class $C^2$, nondecreasing, 
with $\kappa_\e(z)=z$ for $z\geq\e$, $\kappa_\e(z)= \e/2$ for $z\in [0,\e/2]$,
with $|\kappa'_\e(z)|+ \e |\kappa_\e''(z)| \leq C$.
Consider then $a_\e,b_\e$ be defined as $a,b$ with 
$|z|^\gamma$ replaced by $[\kappa_\e(|z|)]^\gamma$. 
Then $a_\e$ is of class $C^2$, with
all its derivatives of order $2$ bounded by $C\e^\gamma$, and $b_\e$
is Lipschitz continuous with Lipschitz constant $C\e^{\gamma}$.

We thus may apply Corollary \ref{corpm} and Theorem \ref{totaldisc}. 
Denote by $(P^\e_t)_{t\geq 0}=(\loi(X^{1,\e}_t))_{t\geq 0}$ a weak solution to 
(\ref{eqa}) with $a_\e$ and $P_0^\e=P_0$. Then we believe that
our results, plus some moment and ellipticity
estimates (uniform in $\e \in (0,1]$), will
give something like $\E[\sup_{[0,T]}|X^{1,n,N,\e}_t-X^{1,\e}_t|^2]\leq 
(n^{-1}+N^{-1})\exp(C_T\e^{2\gamma})$, where
$(X_t^{i,n,N,\e})_{t\geq 0, i=1,..,n}$ solves (\ref{sdenN}) with $a_\e,b_\e$
instead of $a,b$.

\vip

On the other hand, we may apply the techniques introduced in \cite{fg}
to estimate $\cW_2^2(P_t,P^\e_t)$, where 
$(P_t)_{t\geq 0}=(\loi(X^1_t))_{t\geq 0}$ is a weak solution to (\ref{eqa}) 
with $a$ and $P_0$.
We believe that, with a convenient coupling, 
it is possible to obtain something like
$\sup_{[0,T]} \E[|X^{1,\e}_t-X^1_t|^2] \leq C_{T}\e^2$.

\vip

One would thus get
$\sup_{[0,T]}\E[|X^{1,n,N,\e}_t - X^1_t|^2]\leq 
C_T \left(\e^2 + \left(n^{-1}+N^{-1}\right)e^{C_T\e^{2\gamma}}\right)$.
This is of course an awfull rate of convergence.
It does not seem reasonnable to handle a rigorous proof.

\subsection*{Simulation without cutoff}
However, the particle system (\ref{sdenN}) is still well-defined and simulable
for soft potentials (with $\gamma \in [-3,0]$),
at least if we replace $\frac{1}{n}\sum_{k} \delta_{X^{k,n,N}_t}$
by  $\frac{1}{n}\sum_{k\ne i} \delta_{X^{k,n,N}_t}$ and if $P_0$ has a density.
Based on the well-posedness result of \cite{fg}, we hope that, at least when
$\gamma \in (-2,0]$, one might obtain
the same estimates as in Corollary \ref{corpm} and Theorem 
\ref{totaldisc} (
under additionnal conditions on $P_0$).
The proof however seems to be quite difficult:  
we do not know how to get a sufficiently good estimate of
quantities like $|X^{i,n,N}_t-X^{j,n,N}_t|^\gamma$.

\section{Numerics}\label{num}

Let us first observe that for the Landau equation \ref{eqa}
where $a$ is given by (\ref{aLandau}) and $b_i=\sum_j \partial_j a_{ij}$,
the simulable particle system (\ref{sdenN}) is conservative, in the
sense that it preserves, in mean, momentum and kinetic energy:
for all $i=1,\dots,n$, all $t\geq 0$, $\E[X^{i,n,N}_t]=\int x P_0(dx)$
and $\E[|X^{i,n,N}_t|^2]=m_2(P_0)$.

\vip

We consider here the Landau equation for soft potentials,
for some $\gamma \in [-3,0]$,
described in the previous section, in dimension $d=2$. We use no
cutoff procedure in the case $\gamma<0$. We consider the initial
condition $P_0$ with density $P_0(x_1,x_2)=f(x_1)g(x_2)$,
where $f$ is the Gaussian density with mean $0$ and variance $0.1$,
while $g(x)=(f(x-1)+f(x+1))/2$. The momentum and energy of $P_0$ are
given by $(0,0)$ and $1.02$. 

Thus in large time, the solution $P_t$ should converge to the
Gaussian distribution with mean $(0,0)$ and covariance matrix
$0.51 I_2$, see Villani \cite{v3}.

We use the particle
system (\ref{sdenN}) with $n$ particles, and $N$ steps per unit of time.
Easy considerations show
that the computation of  (\ref{sdenN}) until time $T$ is
essentially proportionnal to $TNn^2$, and should not depend too much
on $\gamma$. However, it is consequently faster when $\gamma=0$
for obvious computational reasons.
Let us also remark that the law of (\ref{sdenN})
does not change when replacing $\ah$ by any $\sigma$
such that $\sigma(x,\mu)\sigma(x,\mu)^*=a(x,\mu)$. 
We thus use a Cholesky decomposition, which is numerically quite fast.
Let us give an idea of the time needed to perform one time-step:
with $\gamma=0$, it takes around $7.10^{-3}$ seconds ($n=500$), $0.15$ s
($n=2500$), $3.5$ s ($n=12500$), and $13$ s
($n=25000$). The computations are around $10$ times slower when $\gamma<0$.

\vip

Now we alway use $n=5000$ particles, and $N=200$ steps
per unit of time. We draw, for different values of $t$ and $\gamma$,
the histogram (with $80$ sticks) based on the 
second coordinates of $(X_t^{i,n,N})_{i=1,\dots,n}$.
The plain curve is the expected asymptotic Gaussian density, with mean
$0$ and variance $0.51$.
The convergence to equilibrium seems to be slower and slower as 
$\gamma$ is more and more negative. 

\begin{center}
\includegraphics[width=5cm]{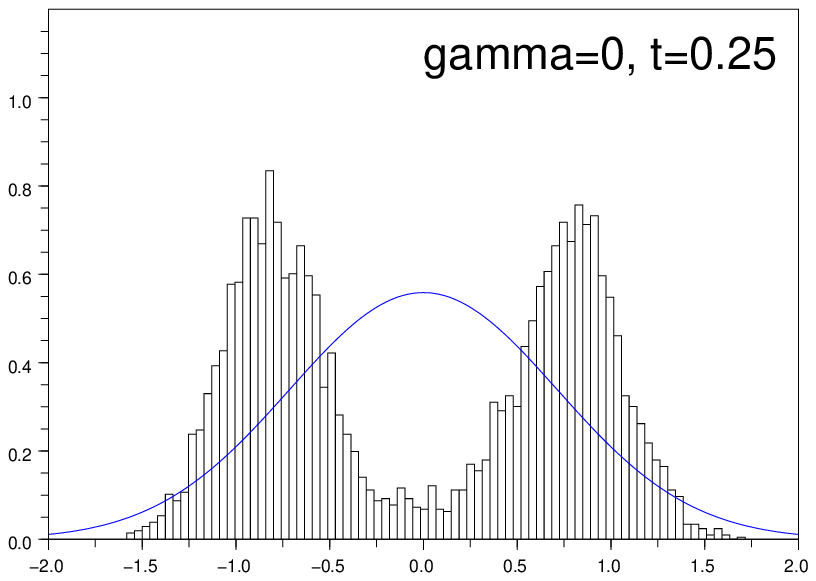}\hskip-0.3cm
\includegraphics[width=5cm]{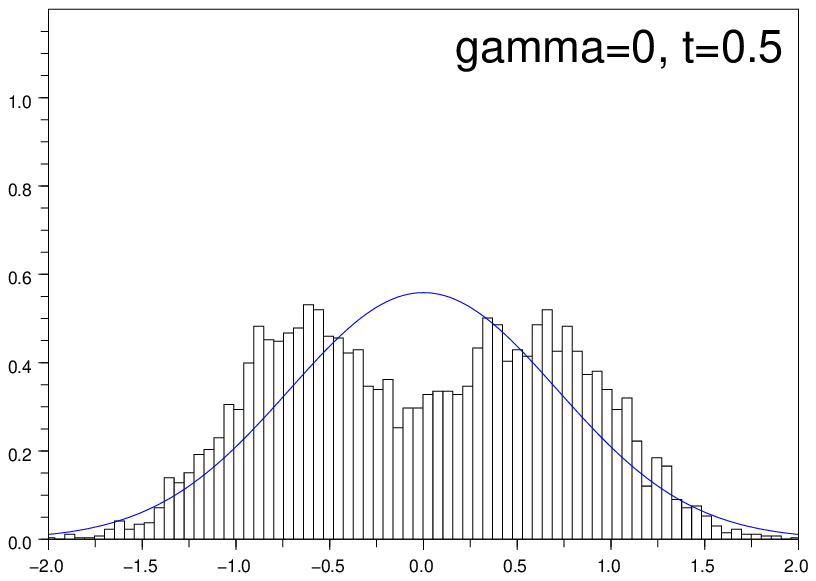}\hskip-0.3cm
\includegraphics[width=5cm]{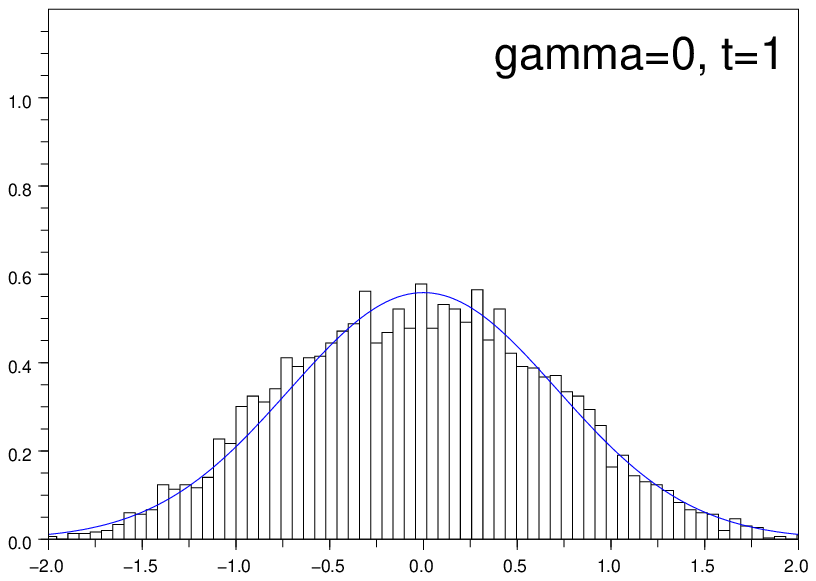}
\end{center}
\begin{center}
\includegraphics[width=5cm]{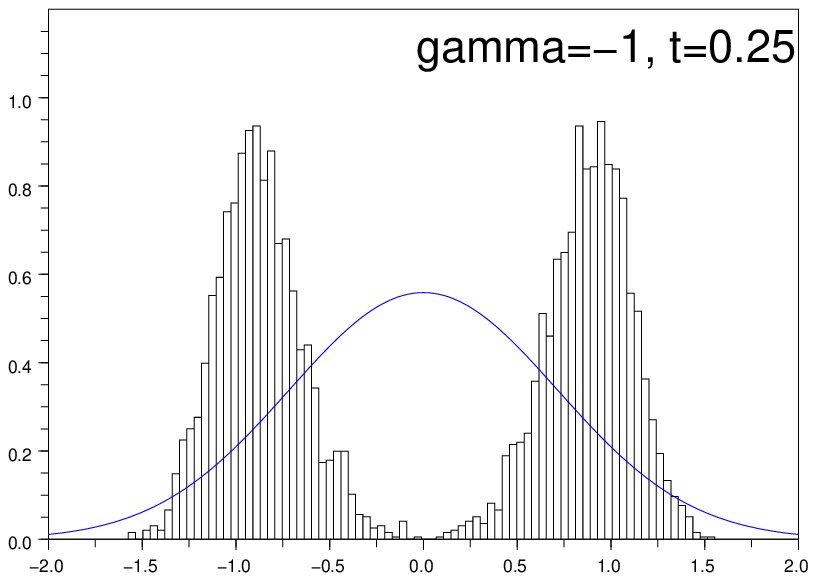}\hskip-0.3cm
\includegraphics[width=5cm]{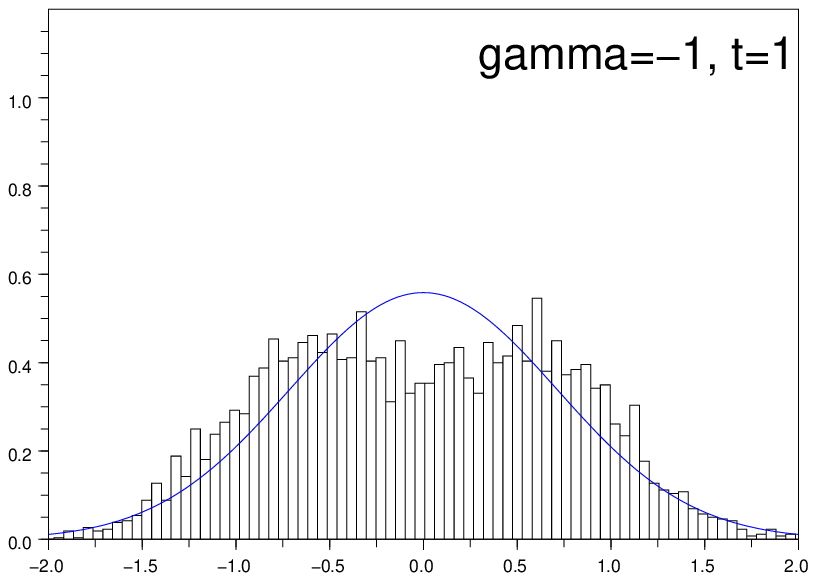}\hskip-0.3cm
\includegraphics[width=5cm]{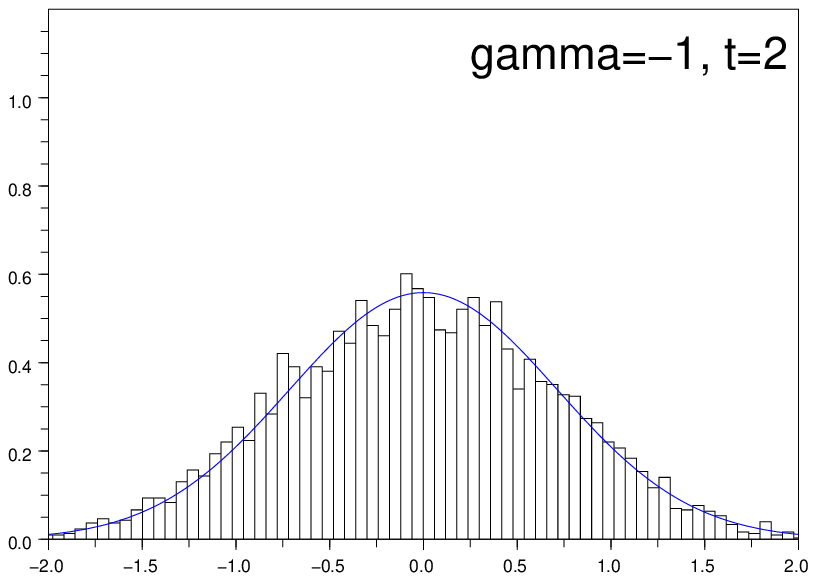}
\end{center}
\begin{center}
\includegraphics[width=5cm]{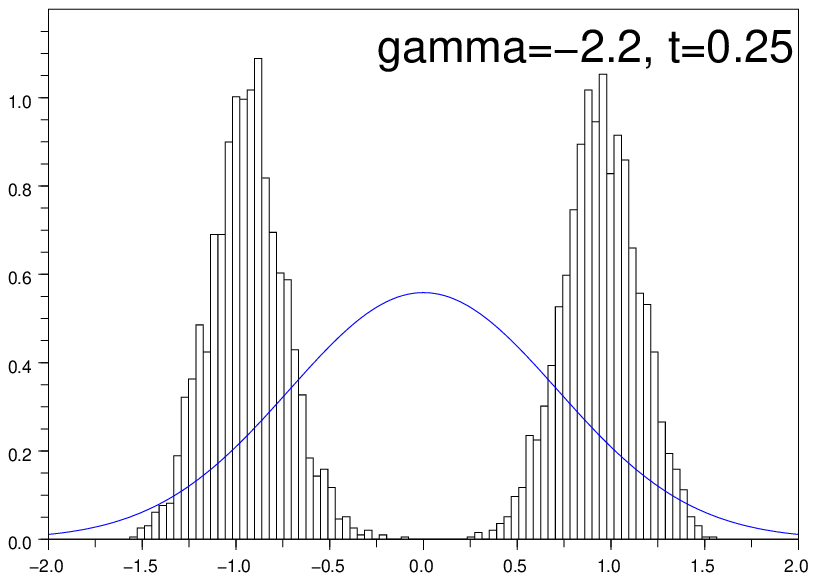}\hskip-0.3cm
\includegraphics[width=5cm]{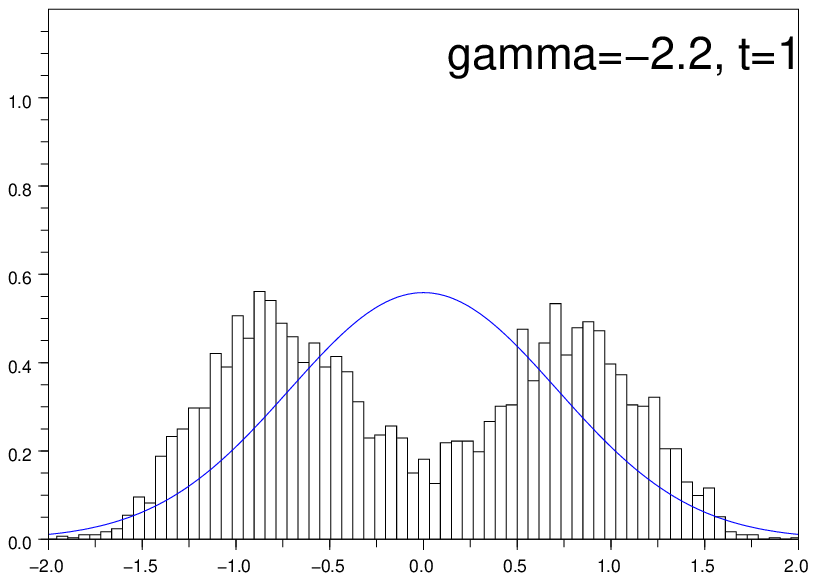}\hskip-0.3cm
\includegraphics[width=5cm]{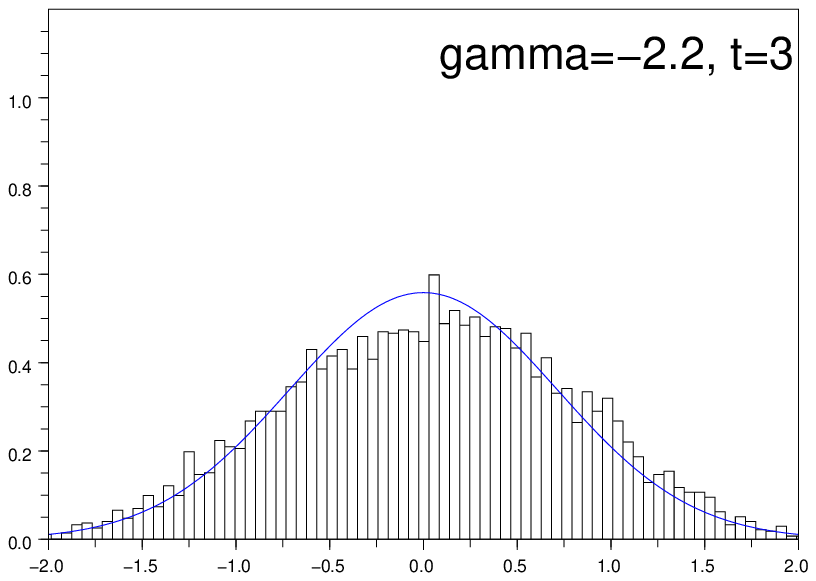}
\end{center}

For too small values of $\gamma$ (say $\gamma<-2.5$), 
the numerical results are not so convincing.
This is not surprising, since the coefficients are more and more singular
as $\gamma$ becomes smaller and smaller.

\section{Appendix}\label{app}

The following Lemma can be found in Stroock-Varadhan (when $p=d$) 
\cite[Theorem 5.2.3]{sv}, or in
Villani \cite[Theorem 1]{v5} (for a more refined statement including
all possible values of $p$ and $d$).

\begin{lem}\label{a1}
Let $A:\rr^p \mapsto S_d^+$, for some $p\geq 1$, $d\geq 1$, be of class $C^2$,
with all its derivatives of order $2$ bounded.
Then $||D (A\sq)||_\infty \leq C_{p,d} \sqrt{ ||D^2 A ||_{\infty}}$, where 
$C_{p,d}$ depends only on $p$ and $d$.
\end{lem}

We also need the following estimates, which are probably standard.

\begin{lem}\label{a1bis} 
For $A,B \in S_d^+$,

(i) there holds $|A\sq-B\sq| \leq \sqrt{|A-B|}$

(ii) and $|A\sq-B\sq| \leq \sqrt{\min(|A^{-1}|,|B^{-1}|)}\times |A-B|$.
\end{lem}

\begin{proof}
We start with point (i). Let
$\sigma=|A\sq-B\sq|$. 
There is a unit vector $e\in\rd$ such that $|(A\sq-B\sq)e|=\sigma e$, and
we may assume that $(A\sq-B\sq)e=\sigma e$ 
(else, change the roles of $A,B$).
Then, using that $B\sq$ is nonnegative, 
\begin{align*}
|A-B|\geq (Ae-Be,e)=((A\sq-B\sq)e,(A\sq+B\sq)e)=
(\sigma e,\sigma e + 2B\sq e)
\geq \sigma^2|e|^2=\sigma^2.
\end{align*}
We now prove (ii). First observe that $(A\sq x,x)\geq |x|^2/ |A^{-1/2}|$
for all $x\in \rd$. As previously,
\begin{align*}
|A-B|\geq& ((A\sq-B\sq)e,(A\sq+B\sq)e)=\sigma (A\sq e,e)+
\sigma(B\sq e,e)\ala
\geq & \sigma |e|^2/ |A^{-1/2}|  + \sigma |e|^2/ |B^{-1/2}| 
= \sigma/\sqrt{|A^{-1}|}+\sigma/\sqrt{|B^{-1}|} 
\geq \sigma/\sqrt{\min(|A^{-1}|,|B^{-1}|)},
\end{align*}
whence $|A\sq-B\sq|=\sigma \leq \sqrt{\min(|A^{-1}|,|B^{-1}|)}|A-B|$.
\end{proof}

We conclude this annex with an elementary fact on the Wasserstein
distance.

\begin{lem}\label{a2}
For $x_1,\dots,x_n$, $y_1,\dots,y_n \in \rd$,
$\cW^2_2\left(\frac{1}{n}\sum_{1}^n\delta_{x_i},\frac{1}{n}\sum_{1}^n\delta_{y_i}
\right) \leq \frac{1}{n}\sum_{1}^n |x_i-y_i|^2$.
\end{lem}

\begin{proof}
Let $U$ be uniformly
distributed on $\{1,\dots,n\}$, set $X=x_U$ an $Y=y_U$. Then 
$X\sim \frac{1}{n}\sum_{1}^n\delta_{x_i}$, 
$Y\sim \frac{1}{n}\sum_{1}^n\delta_{y_i}$, 
and $\E[|X-Y|^2]= \frac{1}{n}\sum_{1}^n |x_i-y_i|^2$.
\end{proof}


\begin{thebibliography}{99}

\bibitem{bcf}{{\sc C. Buet, S. Cordier, F. Filbet}, 
{\it Comparison of numerical schemes for Fokker-Planck-Landau equation.},  
CEMRACS 1999 (Orsay),  161-181, ESAIM Proc., 10,
 Soc. Math. Appl. Indust., Paris, 1999.}

\bibitem{cm}{{\sc S. Cordier, S. Mancini}, {\it A brief review on numerical
methods for the collision operators}, ENUMATHS Proceedings 2001, 
Springer, 2003.}

\bibitem{dv}{{\sc L. Desvillettes, C. Villani,} {\it On the spatially 
homogeneous Landau equation for hard potentials Part I: existence, uniqueness, 
and smoothness}, Comm. Partial differential equations, 25, no 1-2, 
175-259, 2000.}

\bibitem{fgm}{{\sc J. Fontbona, H. Gu\'erin, S. M\'el\'eard},
{\it Measurability of optimal transportation and convergence rate for Landau 
type interacting particle systems}, to appear in
Probab. Theory Related Fields, 2008.}

\bibitem{fg}{{\sc N. Fournier, H. Gu\'erin}, {\it Well-posedness
of the spatially homogeneous Landau equation for soft potentials},
Preprint, 2008.}

\bibitem{f}{{\sc T. Funaki}, {\it The diffusion approximation of the spatially
homogeneous Boltzmann equation}, Duke Math. J. 52, 1-23, 1985.} 


\bibitem{g}{{\sc H. Gu\'erin}, {\it Solving Landau equation for some 
soft potentials through a probabilistic approach}, Ann. Appl. Probab. 13, 
no. 2, 515--539, 2003. }

\bibitem{sv}{{\sc D.W. Stroock, S.R.S. Varadhan}, Multidimensional diffusion
processes, Reprint of the 1997 edition. Classics in Mathematics. Springer, 
2006.}

\bibitem{v5}{{\sc P.L. Lions, C. Villani},
{\it R\'egularit\'e optimale des racines carr\'ees},
C.R. Acad. Sci. 321, 1537-1541, 1995. }

\bibitem{v}{{\sc C. Villani},
{\it On the spatially homogeneous Landau equation for Maxwellian molecules.}, 
Math. Models Methods Appl. Sci. 8, no. 6, 957--983, 1998.}

\bibitem{v2}{{\sc C. Villani}, 
{\it On a new class of weak solutions to 
the spatially homogeneous Boltzmann and Landau equations}, Arch. Rational 
Mech. Anal.  143,  no. 3, 273--307, 1998.}

\bibitem{v3}{{\sc C. Villani}, {\it A review of mathematical topics 
in collisional kinetic theory}, Handbook of mathematical fluid dynamics, 
Vol. I,  71--305, North-Holland, Amsterdam, 2002.}

\bibitem{v4}{ {\sc C. Villani}, {\it Topics in Optimal Transportation,}
Graduate Studies in Mathematics Vol. 58, AMS 2003.}

\bibitem{w}
{{\sc J.B. Walsh,}  {\it An introduction to  stochastic partial
differential equations,} \'Ecole d'\'et\'e de Probabilit\'es
de Saint-Flour XIV, Lect. Notes in Math. {{1180}},
265-437, 1986.}

\end{thebibliography}
\end{document}